**Nonrelaxational FMR peak broadening in spatially inhomogeneous films**

Victor A. L'vov[a,b], Julia Kharlan[b,c,*], Vladimir O. Golub[b]

[a]*Taras Shevchenko National University of Kyiv, Kyiv 01601, Ukraine*

[b]*Institute of Magnetism NASU and MESU, Kyiv 03142, Ukraine*

[c]*Faculty of Physics, ISQI, Adam Mickiewicz University, Poznan 61-614, Poland*

**Abstract**

The modification of magnetic properties in spatially inhomogeneous epitaxial films of magnetic shape memory alloys in martensitic state with the temperature variation has been studied. The proposed theoretical model is based on Landau theory of martensitic transformation and statistical model of martensitic state. It was shown that that spatial inhomogeneity of the material leads to the dispersion of local martensitic transformation temperatures resulting in the variation of local magnetic anisotropy values. This model allows describing the dramatic ferromagnetic resonance line broadening observed in the experiments in epitaxial films of magnetic shape memory alloys at low temperatures.

*Keywords: ferromagnetic resonance, spatial inhomogeneities, resonance line broadening, magnetic shape memory alloys, martensitic transformation*

**1. Introduction**

Spatial inhomogeneity is frequently encountered feature of ferromagnetic films (see for instance [1]). The spatial inhomogeneity is inherent, in particular, to the films of magnetic shape memory alloys (MSMAs) [2, 3]. For MSMAs it can be in particular related to the hierarchic microstructure arising due to the structural martensitic transformation (MT) from the high-temperature cubic phase (austenite) to the low-temperature phase with lower symmetry of crystal lattice (see e.g. [4])). The experimental and theoretical study of magnetic properties of MSMA films is motivated by the potential applications of MSMAs ranging from magnetic nanoelectronics to magnetic refrigeration and magnetic actuators [5-8]. Most of these applications require an extensive knowledge about the relation of magnetic and magnetodynamic parameters of MSMA films with their spatial and electronic structure. That is why the studies in this area continue to grow [5]).

---

[*] Corresponding author: *julia-lia-a@ukr.net*

Ferromagnetic resonance (FMR) is one of the most powerful technique for the investigation of magnetic properties the materials, which provides information about their magnetization and magnetic anisotropy, magnetic phases, magnetostatic, magnetoelastic and exchange coupling, magnetodynamics, mechanisms of magnetic relaxation, etc. [9]. A lot of experiment in particular is devoted to the study of the resonance linewidth to obtain the information about microwave energy dissipation from spin system to lattice (homogeneous broadening), which is critical, for instance, for magnonic applications, and about different magnetic imperfections (inhomogeneous broadening). Magnetic relaxation rate and FMR linewidth depend on the microwave frequency, temperature of ferromagnetic specimen and some other factors, including the microstructure and defect structure of real specimens (see e. g. [10] and references therein). Normally, FMR lines broaden on heating of ferromagnetic solid and their width most strongly increases when the temperature approaches Curie temperature [9, 11, 12], due to the thermal fluctuations and the rapid decrease of the magnetization value. But the FMR measurements of MSMAs films in martensitic state showed the peculiar feature, which is not typical for the most others ferromagnets: a drastic increase of the resonance linewidth with the temperature decrease. For polycrystalline films, this increase was more or less successfully explained in terms of the increase of magnetocrystalline and magnetoelastic anisotropy value below the MT temperature due to the random orientation of magnetic anisotropy axis (see, for instance, [13]). However, it has been shown recently (see, for example [14]) that a strong increase of resonance linewidth with the temperature increase can be observed in epitaxial films, where practically perfect alignment of magnetic anisotropy axis takes place. These results have clearly demonstrate that the situation is much more complicated and another approaches to understand the resonance line broadening should be developed. Taking into account strong variation of FMR linewidth in the temperature range where the variation of the magnetization is negligible it can be suggested that the observed broadening of FMR peaks is not related to the temperature dependence of the relaxation rate and can be referred to the nonrelaxational broadening (NB).

Basic characteristics of MT for bulk MSMAs and their thin films strongly depend on their chemical composition, defect structure, atomic ordering, internal mechanical stress, etc. All these factors are different for different spatial domains of MSMAs. This fact had motivated the development of the statistical model of martensite [15]. This model treats the spatially inhomogeneous martensitic state as a statistical ensemble of small spatial domains with different deformational properties [15] and Curie temperatures [16].

Here we present the consideration of FMR in martensitic film, which is based on the combination of the statistical model of the spatially inhomogeneous martensitic state with the well-elaborated Landau theory of cubic-tetragonal MTs observed in the widely studied Ni-Mn-Ga alloys (analytic survey of the Landau-type theories of MTs is presented in the recent publication [17]). Although in real martensitic films the symmetry of martensitic phase can be lower (orthorhombic or even monoclinic) the theoretical results, obtained for cubic-tetragonal MT, provide a satisfactory description of the observed temperature dependence of FMR peak width. The consideration presented below is mainly focused on the analysis of FMR linewidth but the developed approach to the description of the resonance properties of martensitic film can be used for the analysis of other magnetic data (magnetic susceptibility, magnetic hysteresis loops, etc.). It should be also pointed out that this work gives a clue for understanding of complex temperature behavior of net magnetic anisotropy of real MSMA samples (see e.g. [18]).

## 2. Model formulation

The proposed model is based on the following suppositions: i) spatial inhomogeneity of MSMA exerts influence on the martensite start and martensite finish temperatures; ii) lattice parameters in the martensitic phase strongly depend on the temperature (for relevant experimental data see for example [19], [20]). According to these suppositions a spatially inhomogeneous martensitic film can be presented as an ensemble of $N$ spatially homogeneous domains enumerated by the indexes $i = 1, 2, 3 ... N$. Each domain is characterized by the martensite start and martensite finish temperatures $T_{MS}^{(i)}$ and $T_{MF}^{(i)}$. The temperatures are enumerated in the ascending order. According to the statistical model of martensite [15], the volume fractions of domains are assumed to be described by the normal distribution

$$p^{(i)} = \frac{1}{Z} \exp\left\{-\frac{[T_{MS}^{(i)} - T_{MS}^{(N/2)}]^2}{2T_G^2}\right\},$$
$$Z = \sum_{i=1}^{N} \exp\left\{-\frac{[T_{MS}^{(i)} - T_{MS}^{(N/2)}]^2}{2T_G^2}\right\},$$
(1)

where $T_G$ is Gaussian root square width. Index $i = N/2$ corresponds to the domain with maximal volume fraction, having the average value of MT start and MT finish temperatures $T_{MS}^{(N/2)} \equiv T_{MS}$, $T_{MF}^{(N/2)} \equiv T_{MF}$. For the spatially inhomogeneous film not only the martensite start and martensite finish temperatures, but also the difference of "local" MT temperatures, $\Delta T^{(i)} \equiv T_{MS}^{(i)} - T_{MF}^{(i)}$ have to be dependent on the index $i$. It is advantageous to express this difference as

$$\Delta T^{(i)} = \begin{cases} \Delta T^{(N/2)} + \psi[T_{MS}^{(N/2)} - T_{MS}^{(i)}], & \text{if } T_{MS}^{(i)} > T_{MF}^{(i)} + 1, \\ \Delta T^{(i)} = 1 \text{ K}, & \text{otherwise.} \end{cases}$$
(2)

Equation (2) takes into account that minimal temperature intervals of experimentally observed MTs are of about 1 K (see [21, 22] and references therein). The dimensionless parameter $\psi = -d\Delta T^{(i)}/dT_{MS}^{(i)}$ is positive if the factors, which retard the start of MT, result in the broadening of the temperature interval of MT; the negative values of parameter $\psi$ correspond to the case of narrowing of the MT temperature interval; zero value of this parameter corresponds to the temperature interval of MT for a spatially homogeneous alloy, $\Delta T^{(i)} = \Delta T^{(N/2)}$.

The "tetragonality" of crystal lattice in the domain $i$ of martensitic phase can be characterized by the parameter $(1-c/a)^{(i)}$, where $a$ and $c$ are the lattice parameters in $i$-th domain at the temperature $T$. According to Landau theory of cubic-tetragonal MT the lattice tetragonality is related to MT temperatures as

$$\left(1-\frac{c}{a}\right)^{(i)} = \frac{1}{2}\left(1-\frac{c}{a}\right)^{(N/2)}_{MF} \left\{1+\left[\frac{T_{MS}^{(i)}-T}{T_{MS}^{(i)}-T_{MF}^{(i)}}\right]^{1/2}\right\}, \qquad (3)$$

where $(1-c/a)^{(N/2)}_{MF}$ is the tetragonality at the martensite finish temperature for the domain with $i = N/2$ (see the Appendix, for more details). Equation (3) and numerous experiments (see for example [19,20]) show that the absolute value of lattice distortion substantially increases when MSMAs is cooled down from martensite start to liquid helium temperatures. As it is known, the magnetocrystalline anisotropy of ferromagnetic solid is characterized by the magnetic anisotropy constant. The uniaxial magnetocrystalline anisotropy $K_u$ of small spatial domains of tetragonal martensitic phase is datermined by

$$K_u^{(i)}(T) = -6\delta(1-c/a)^{(i)} M^2(T), \qquad (4)$$

where $M(T)$ is the saturation magnetization of MSMA, $\delta$ is the dimensionless constant (see Ref. [23] and references therein). These parameters are temperature-dependent because of the temperature dependence of the tetragonality and the magnetization. Magnetization values $M^{(i)}(T)$ are different for different domains, but this difference is significant only near Curie temperature and can be disregarded for temperatures much below the Curie temperature.

The resonance magnetic field for perpendicular magnetized film can be found using the standard equation

$$H_{\text{res}}^{(i)} = \frac{\omega}{\gamma} + 4\pi M(T) + \frac{K_u^{(i)}(T)}{M(T)}, \tag{5}$$

where $\omega$ is the resonance frequency and $\gamma$ is the gyromagnetic ratio. The microwave absorption of each martensitic domain of the film can be considered to be proportional to the volume fraction $p^{(i)}$ of this domain in martensitic phase. The experiments show that the "residual" austenitic phase is present in FSMAs well below the martensite finish temperature. Therefore, the microwave power absorption in domain $i$ can be described as

$$I^{(i)}(T) = \alpha(T) p^{(i)}, \tag{6}$$

where

$$\alpha(T) = \frac{1}{2}\left[1 + \tanh\left(\frac{T_0 - T}{\Delta}\right)\right], \tag{7}$$

is the volume fraction of martensite in the film, which is equal to 0.5 at the temperature $T_0 = (T_{MS} + T_{MF})/2$ and becomes close to unit well below the average value of the martensite finish temperature; parameter $\Delta = (T_{MS} - T_{MF})/2$ is the half-width of the temperature interval of MT. Equations (5) – (7) enable to calculate a large number of coupled values $(H_{\text{res}}^{(i)}, I^{(i)})$ and to plot the dependences of microwave power absorption $I(H,T)$ and its derivative ($\partial I / \partial H$) for some definite temperature. The interval between the extremums of this derivative is the FMR linewidth usually measured in experiments.

## 3. Computations and results

The abnormal temperature dependence of the linewidth is quite common for martensitic films, but here we present the comparison of the theoretical results with experimental data for Ni-Mn-Ga film taken from Ref. [14].

The experimental temperature dependences of the FMR linewidth, the magnetization value and the resonance field value are shown in Fig. 1. Figure 1 (a) illustrates a pronounced

FMR-peaks broadening with the temperature decrease. The temperature dependence of the magnetization (Fig. 1 (b)) was fitted by the function $M(T)$, which was used for computation of the resonance field values. Figure 1 (c) shows the experimental values of the resonance field and the average resonance field $H_{res}^{(N/2)}(T)$ computed using the temperature dependent "tetragonality" of crystal cells (see Fig. 2) and magnetoelastic constant $\delta = -23$ evaluated formerly for the Ni-Mn-Ga alloys [23]. The MT temperature 420 K was reported for this film in Ref. [14]. This temperature was treated as the average value of the martensite start temperature $T_{MS}^{(N/2)}$ for the computation of the temperature-dependent parameter $(1-c/a)^{(N/2)}$, required for the determination of $K_u^{(N/2)}(T)$ and $H_{res}^{(N/2)}(T)$. The value $(1-c/a)_{MF}^{(N/2)} = -0.0225$ was adjusted to fit the resonance field value computed for $T = 150$ K to experimental value measured for this temperature (see Fig. 1 (c)).

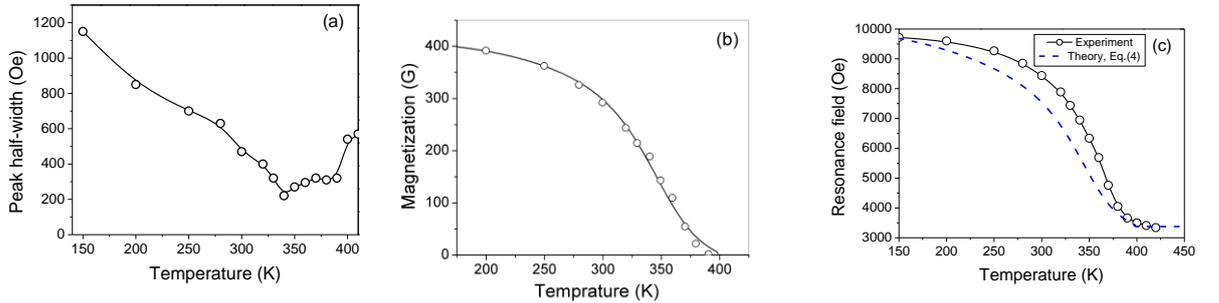

**Fig. 1.** Experimental temperature dependence of the peak width [14], (a); experimental temperature dependence of magnetization (circles) and plot of $M(T)$ function, used for computations, (b); temperature dependence of the resonance field, (c).

Figure 1 shows that theoretical $H_{res}^{(N/2)}(T)$ curve is reasonably close to the experimental points. The deviation of the curve from experimental points is expectable because the symmetry of crystal lattice in the martensitic film [14], is lower than tetragonal and therefore, the temperature dependence of its magnetocrystalline anisotropy parameter cannot be precisely described by the Eq. (4).

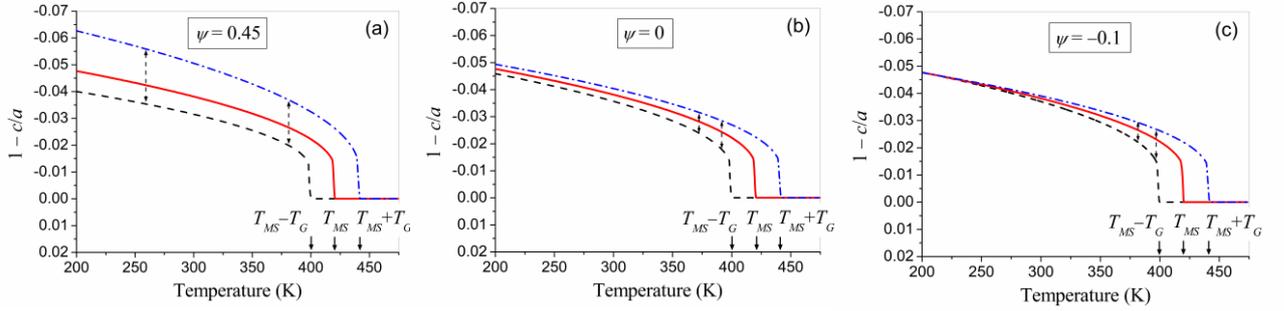

**Fig. 2.** Theoretical temperature dependencies of the tetragonality of unit cells of crystal lattice in the domains with MT start temperatures shown by the one-side arrows; $T_{MS}$ is the average value of the local martensite start temperatures; two-side arrows show the increase, (a), and the decrease, (b), (c) of the difference between tetragonality values of the domains

Fig. 2 shows the dependence of the tetragonality of unit cells on the temperature. The tetragonality values were computed using Eqs. (1) – (3) with $\Delta T = T_G = 20$ K for different values of parameter $\psi$. The two-side arrows in Fig. 2 show that MSMA cooling down results in in the increase of the difference in the tetragonality values in the spatial domains with different $T_{MS}$ if a decrease of MT temperature is accompanied by the broadening of the temperature interval of MT ($\psi = 0.45$). In this case the broadening of FMR peak should take place. This conclusion is confirmed by the computations performed for the film experimentally studied in [14]: the reduction of MT start temperature and broadening of the temperature interval of MT ($\psi > 0.1$) results in the broadening of FMR peak from 190 Oe to 870 Oe when the film is cooling down from 350 K to 150 K (see Fig. 3). If the reduction of MT start temperature is accompanied by the narrowing of the temperature interval of MT, the peak width becomes a non-monotonic function of the temperature, which tends to zero with the temperature decrease.

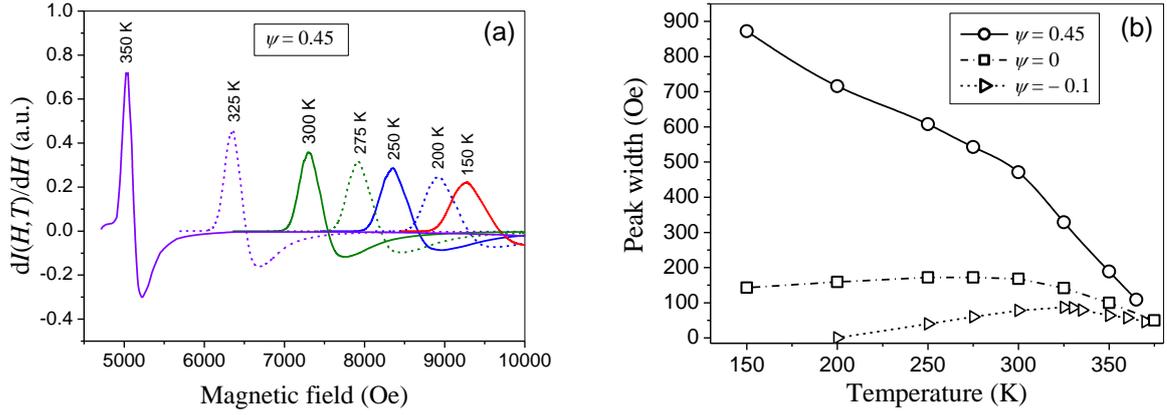

**Fig. 3.** Theoretical calculations FMR signal parameters: (a) the peak profiles, computed for $\psi$=0.45 and different temperatures, (b); the results of computations are shown by symbols. The lines are the guides for eye.

## 4. Conclusion

The statistical model of the spatially inhomogeneous martensitic state and Landau theory of cubic-tetragonal MTs were adopted for the analysis of magnetic anisotropy and ferromagnetic resonance of MSMAs below the MT point. It has been shown that spatial inhomogeneity of real MSMAs (in particular, MSMA films) leading to a difference of "local" MT temperatures can result in "local" variation of magnetic anisotropy constants. The presented statistical model allowed describing dramatic broadening of the FMR line observed for MSMAs epitaxial films with the temperature decrease. It is worth to be noted, that spatial inhomogeneity of hierarchically structured martensitic states can result in nontrivial temperature behavior of the net magnetic anisotropy of MSMAs.

## 5. Appendix: cubic-tetragonal MT in nonideal crystal lattice

Landau theory describing the phase transformation of cubic cell of the crystal into tetragonal cell with the four-fold symmetry axis parallel to $z$-axis of coordinate frame was described using the following expression for Gibbs free energy of deformable cubic crystal:

$$G_{el} = \frac{1}{2}c_1 u_1^2 + \frac{1}{2}c_2(T)u_3^2 + a_2 u_1 u_3^2 + \frac{1}{3}a_4 u_3^3 + \frac{1}{4}b_4 u_3^4 + \frac{1}{2}b_7 u_1 u_3^3 + PV, \qquad (A.1)$$

where the variable $u_1$ is related to the relative volume change of the crystal lattice, $v = \Delta V / V$, as $u_1 = v/3$, the variable $u_3 = 2\varepsilon_{zz} - \varepsilon_{yy} - \varepsilon_{xx}$, $\varepsilon_{ik}$ are the strain tensor components, coefficients $c_1$, $c_2(T)$, $a_2$, $a_4$, and $b_4$ are the phenomenological constants [24]. These constants are the linear combinations of second-, third-, and fourth-order elastic modules.

For the spontaneous strain arising as a result of cubic-tetragonal MT of the ideal crystal the strain tensor components satisfy the condition $\varepsilon_{xx} = \varepsilon_{yy} \approx -\varepsilon_{zz}/2$, and therefore,

$$|v| << |\varepsilon_{zz}|, \; u_3 \approx 3\varepsilon_{zz}. \qquad (A.2)$$

For the martensitic films the volume change accompanying MT can be much larger then for the single crystals [25]. One can describe the "additional" volume change $v_{ad}$ introducing the "internal" pressure $P_{in} = -B v_{ad}$, where $B$ is bulk modulus. Internal pressure is negative, if there is an internal factor, which prevents the close packing of atoms.

The condition $\partial G_{el} / \partial u_1 = 0$ results in the direct proportionality of variable $u_1$ to $P_{in} u_3^2$. Due to this, the free energy Eq. (A.1) can be presented in the form

$$G_{el} = \frac{1}{2}c_2^*(T)u_3^2 + \frac{1}{3}a_4^*(P)u_3^3 + \frac{1}{4}b_4^* u_3^4, \qquad (A.3)$$

where

$$c_2^*(T,P) = c_2(T) - 3a_2 P_{in}/c_1,$$

$$a_4^*(P) = a_4 - 9b_7 P/2c_1, \qquad (A.4)$$

$$b_4^* = b_4 - a_2^2/2c_1.$$

The spontaneous strain arising as a result of cubic-tetragonal MT satisfies condition $\partial G_{el}/\partial u_3 = 0$, which results in the relationship

$$u_3(T, P_{in}) = -\frac{a_4^*(P_{in})}{2b_4^*}\left\{1 + [1 - c_2^*(T, P_{in})/c_t(P_{in})]^{1/2}\right\}, \quad (A.5)$$

where $c_t(P_{in}) = a_4^{*2}(P_{in})/4b_4^*$.

MT starts at

$$c_2^*(T, P_{in}) = c_t(P_{in}) \quad (A.6)$$

and finishes at

$$c_2^*(T, P_{in}) = 0. \quad (A.7)$$

It is important that Eq. (A.6) for MT start temperature involves the internal pressure multiplied by factors $a_2/c_1$ and $b_7/c_1$, while Eq. (A.7) for MT finish temperature involves $P_{in}$ multiplied by $a_2/c_1$ only. Due to this, the influence of crystal lattice imperfection on MT start temperature may be different from its influence on MT finish temperature. According to the Landau theory of phase transition the coefficient $c_2^*(T, P_{in})$ is assumed to be the linear function of temperature. The equations (A.6), (A.7) have the solution

$$c_2^*(T, P_{in}) = c_t \frac{T - T_{MF}(P_{in})}{T_{MS}(P_{in}) - T_{MF}(P_{in})}. \quad (A.8)$$

Taking into account Eqs. (A.2), (A.8) and equation

$$u_3(T_{MF}, P_{in}) = a_4^*(P_{in})/b_4^*, \quad (A.9)$$

which follows from Eqs. (A.5), (A.6), one can interrelate the lattice parameters of tetragonal phase:

$$1 - \frac{c}{a} = \frac{1}{2}\left(1 - \frac{c}{a}\right)_{MF}\left\{1 + \left[\frac{T_{MS}(P_{in}) - T}{T_{MS}(P_{in}) - T_{MF}(P_{in})}\right]^{1/2}\right\}. \quad (A.10)$$

The lattice parameters of tetragonal crystal lattice are related to spontaneous strain $\varepsilon_{zz}(T)$ arising as a result of MT as $1 - c/a = 3\varepsilon_{zz}/2$.

**Acknowledgements**

The work was partially supported by National Research Foundation of Ukraine trough the grant #2020.02/0261. JK acknowledges the support from the National Science Center – Poland, the grant No. 2021/43/I/ST3/00550 and POLS grant No 2020/37/K/ST3/02450. VOG is grateful for the support from National Academy of Sciences of Ukraine through the grant 0123U100898.